\documentclass[12pt]{article}
\usepackage{graphicx} 
\usepackage{amsmath}
\usepackage{amsfonts}   
\usepackage{amssymb}

\def\om{\omega}
\def\ph{\phi}

\def\p{\partial}
\def\no{\nonumber}
\begin{document}
\date{}









\title{{\bf{\Large Noncommutative Black Hole Thermodynamics }}}
\author{
 {\bf {\normalsize Rabin Banerjee}$
$\thanks{E-mail: rabin@bose.res.in}},\, 
 {\bf {\normalsize Bibhas Ranjan Majhi}$
$\thanks{E-mail: bibhas@bose.res.in}} {\bf {\normalsize and}} 
{\bf {\normalsize Saurav Samanta}
\thanks{E-mail: saurav@bose.res.in}}\\
 {\normalsize S.~N.~Bose National Centre for Basic Sciences,}
\\{\normalsize JD Block, Sector III, Salt Lake, Kolkata-700098, India}
\\[0.3cm]
}

\maketitle

{\bf Abstract:}

We give a general derivation, for any static spherically symmetric metric, of the relation $T_h=\frac{\cal K}{2\pi}$ connecting the black hole temperature ($T_h$) with the surface gravity ($\cal K$), following the tunneling interpretation of Hawking radiation. This derivation is valid even beyond the semi classical regime i. e. when quantum effects are not negligible. The formalism is then applied to a spherically symmetric, stationary noncommutative Schwarzschild space time. The effects of back reaction are also included. For such a black hole the Hawking temperature is computed in a closed form. A graphical analysis reveals interesting features regarding the variation of the Hawking temperature (including corrections due to noncommutativity and back reaction) with the small radius of the black hole. The entropy and tunneling rate valid for the leading order in the noncommutative parameter are calculated. We also show that the noncommutative Bekenstein-Hawking area law has the same functional form as the usual one.

\section{Introduction}
Classical general relativity gives the concept of black hole from which nothing can escape. This picture was changed dramatically when Hawking\cite{hawking} incorporated the quantum nature into this classical problem. In fact he showed that black hole radiates a spectrum of particles which is quite analogous with a thermal black body radiation. Thus Hawking radiation emerges as a nontrivial consequence of combining gravity and quantum mechanics.

After his original derivation which was based on the calculation of Bogoliubov coefficients in the asymptotic states, Hawking together with Hartle\cite{hartle} gave a simpler, path integral derivation. Physically, black hole radiation can be interpreted as the quantum tunneling of vacuum fluctuations through the horizon. This picture was mathematically formulated in\cite{Wilczek}. An important step of this method is the calculation of tunneling amplitude from which the Hawking temperature is obtained. This is done either by using the trajectory of a null geodesic\cite{Wilczek} or by solving the Hamilton-Jacobi equation to calculate the imaginary part of the action variable\cite{Paddy}.

The tunneling approach was subsequently used to compute the Hawking temperature for black holes with different types of metric\cite{Jiang}. The results have agreed with the general formula for the temperature $T_h=\frac{\cal K}{2\pi}$, where $\cal K$ is the surface gravity, a fact that has also been observed in \cite{pilling,sarkar}.

In this paper we derive the relation $T_h=\frac{\cal K}{2\pi}$ using the tunneling approach, thereby making redundant independent calculations of the Hawking temperature for different metrics \cite{Jiang}. By including the self gravitation effect, the imaginary part of the action is computed. Next, exploiting the relation between this and the change in black hole entropy, the Hawking temperature is computed. Interestingly, the derivation is independent of any particular model and true even when quantum effects are incorporated in the metric. The only condition is that the metric should be static and spherically symmetric.

We have applied our general formulation to discuss various thermodynamic properties of a black hole defined in a noncommutative Schwarzschild space time where back reaction is also taken into account. In particular we are interested in the black hole temperature when the radius is very small. Such a study is relevant because noncommutativity is supposed to remove the so called Hawking paradox where for a standard black hole, temperature diverges as the radius shrinks to zero. The Hawking temperature is obtained in a closed form that includes corrections due to noncommutativity and back reaction. These corrections are such that, in some examples, the Hawking paradox is avoided. Expressions for the entropy and tunneling rate are also found for the leading order in the noncommutative parameter.  Furthermore, in the absence of back reaction, we show that the entropy and area are algebraically related in the same manner as occurs in the standard Bekenstein-Hawking area law.   

Before proceeding further, let us mention the organization of the present paper. In the second section we give a general derivation of Hawking temperature in terms of surface gravity. We introduce the noncommutative Schwarzschild space time in the next section. In section 4 we present a thorough analysis of the various thermodynamic entities of a black hole with this metric. The competing roles of noncommutativity and back reaction are illustrated by a graphical analysis. The noncommutative deformation of the Bekenstein-Hawking area law is obtained. The final section is for conclusions after which a short appendix is added. 

\section{Derivation of $T_h=\frac{{\cal{K}}}{2\pi}$}

Calculations of the Hawking temperature, based on the tunneling formalism, for different black holes conform to the general formula $T_h=\frac{\cal K}{2\pi}$. This relation is usually understood as a consequence of the mapping of the second law of black hole thermodynamics ${\textrm d}M=\frac{\cal K}{8\pi}{\textrm d}A$ with ${\textrm d}E=T_h{\textrm d}S_{\textrm{bh}}$, coupled with the Bekenstein-Hawking area law $S_{\textrm{bh}}=\frac{A}{4}$.

Using the tunneling approach, we now present a derivation of $T_h=\frac{\cal K}{2\pi}$ where neither the second law of black hole thermodynamics nor the area law are required. In this sense our analysis is general, going beyond the semiclassical approximation.
 
Consider a metric of the form 
\begin{eqnarray}
ds^2 = -f(r)dt^2+\frac{dr^2}{g(r)}+r^2 d\Omega^2
\label{1.01}
\end{eqnarray}
which describes a general class of static, spherically symmetric space time. There is a coordinate singularity in this metric at the horizon $r=r_h$ where $f(r_h)=g(r_h)=0$. This singularity is avoided by the use of Painleve coordinate transformation\cite{Painleve}, 
\begin{eqnarray}
dt\to dt-\sqrt{\frac{1-g(r)}{f(r)g(r)}}dr. 
\end{eqnarray}
Under this transformation, the metric (\ref{1.01}) takes the following form,
\begin{eqnarray}
ds^2 = -f(r)dt^2+2f(r)\sqrt{\frac{1-g(r)}{f(r)g(r)}} dt dr+dr^2+r^2d\Omega^2.
\label{1.02}
\end{eqnarray}
Note that the metric (\ref{1.01}) looks both stationary and static, whereas the transformed metric (\ref{1.02}) is stationary but not static which reflects the correct nature of the space time. The radial null geodesics are obtained by setting $ds^2=d\Omega^2=0$ in (\ref{1.02}),
\begin{eqnarray}
\dot{r}\equiv\frac{dr}{dt}=\sqrt{\frac{f(r)}{g(r)}}\Big(\pm 1-\sqrt{1-g(r)}\Big)
\label{1.03}
\end{eqnarray}
where the positive (negative) sign gives outgoing (incoming) radial geodesics. At the neighbourhood of the black hole horizon, the trajectory (\ref{1.03}) of an outgoing shell is written as,
\begin{eqnarray}
\dot{r}=\frac{1}{2}\sqrt{f'(r_h)g'(r_h)}(r-r_h)+{\mathcal O}((r-r_h)^2)
\label{1.03a}
\end{eqnarray}
where we have expanded the functions $f(r)$ and $g(r)$ in a Taylor series about $r_h$ 
\begin{eqnarray}
&&f(r)=f'(r_h)(r-r_h)+{\mathcal O}((r-r_h)^2)
\label{1.044}\\
&&g(r)=g'(r_h)(r-r_h)+{\mathcal O}((r-r_h)^2)
\label{1.04}
\end{eqnarray}
and kept the expression of $\dot{r}$ (\ref{1.03}) upto first order only. Now we want to write (\ref{1.03a}) in terms of the surface gravity of the black hole. The reason is that in some cases, for example in the presence of back reaction, one may not know the exact form of the metric but what one usually knows is the surface gravity of the problem. Also, the Hawking temperature is eventually expressed in terms of the surface gravity. The form of surface gravity for the transformed metric (\ref{1.02}) at the horizon is given by, 
\begin{eqnarray}
{\cal{K}}(M)=\Gamma{^0}{_{00}}|_{r=r_h}=\frac{1}{2}\Big[\sqrt{\frac{1-g(r)}{f(r)g(r)}}g(r)\frac{df(r)}{dr}\Big]|_{r=r_h}.
\label{1.05}
\end{eqnarray}
Using the Taylor series (\ref{1.044}) and (\ref{1.04}), the above equation is written as,
\begin{eqnarray}
{\cal{K}}(M)\simeq\frac{1}{2}\sqrt{f'(r_h)g'(r_h)}.
\label{1.06}
\end{eqnarray}
This expression of surface gravity is used to write (\ref{1.03a}) in the form,
\begin{eqnarray}
\dot{r}={\cal{K}}(M)(r-r_h)+{\mathcal O}((r-r_h)^2).
\label{1.07}
\end{eqnarray}

We consider a positive energy shell which crosses the horizon in the outward direction from $r_{{\textrm{in}}}$ to $r_{{\textrm{out}}}$. The imaginary part of the action for that shell is given by
 \begin{eqnarray}
\textrm{Im}~ {\cal{S}} =\textrm{Im} \int_{r_{{\textrm{in}}}}^{r_{{\textrm{out}}}} p_r dr = \textrm{Im}  \int_{r_{in}}^{r_{out}} \int_{0}^{p_r} dp_r'dr.
\label{1.1bb}
\end{eqnarray}
Using the Hamilton's equation of motion $\dot{r}=\frac{dH}{dp_r}|_r$ the last equality of the above equation is written as,
\begin{eqnarray}
\textrm{Im}~ {\cal{S}} 
&=& \textrm{Im} \int_{r_{{\textrm{in}}}}^{r_{{\textrm{out}}}} \int_{0}^{H} \frac{dH'}{\dot{r}} dr
\label{1.1}
\end{eqnarray}
where, instead of momentum, energy is used as the variable of integration.

Now we consider the self gravitation effect\cite{Kraus} of the particle itself, for which (\ref{1.07}) and (\ref{1.1}) will be modified. Following \cite{Wilczek}, under the $s$- wave approximation, we make the replacement $M\rightarrow M-\omega$ in (\ref{1.07}) to get the following expression
\begin{equation}
\dot{r}=(r-r_h){\cal{K}}(M-\omega)
\label{1.3}
\end{equation}
where $\omega$ is the energy of a shell moving along the geodesic of spacetime.

Now we use the fact\cite{Wilczek} , for a black hole of mass $M$, the Hamiltonian $H=M-\omega$. Inserting in (\ref{1.1}) the modified expression due to the self gravitation effect is obtained as,
\begin{eqnarray}
\textrm{Im} ~{\cal{S}} = \textrm{Im} \int_{r_{{\textrm{in}}}}^{r_{{\textrm{out}}}} \int_{M}^{M-\omega} \frac{d(M-\omega)}{\dot{r}} dr  =-\textrm{Im} \int_{r_{{\textrm{in}}}}^{r_{{\textrm{out}}}} \int_{0}^{\omega} \frac{d\omega'}{\dot{r}} dr
\label{1.2}
\end{eqnarray}
where in the final step we have changed the integration variable from $H'$ to $\omega'$. Substituting the expression of $\dot r$ from (\ref{1.3}) into (\ref{1.2}) we find,
\begin{eqnarray}
\textrm{Im} ~{\cal{S}} =  -\textrm{Im} \int_{0}^{\omega} \frac{d\omega'}{{\cal{K}}(M-\omega')}\int_{r_{{\textrm{in}}}}^{r_{{\textrm{out}}}}\frac{dr}{r-r_h}.
\label{1.4cc}
\end{eqnarray}
The $r$-integration is done by deforming the contour. Ensuring that the positive energy solutions decay in time (i.e. into the lower half of $\omega'$ plane and $r_{{\textrm{in}}}>r_{{\textrm{out}}}$) we have after $r$ integration{\footnote{One can also take the contour in the upper half plane with the replacement $M\rightarrow M+\omega$ \cite{Kraus}.}}, 
\begin{eqnarray}
\textrm{Im}~ {\cal{S}}= \pi\int_{0}^{\omega} \frac{d\omega'}{{\cal{K}}(M-\omega')}. 
\label{1.5}
\end{eqnarray}
The tunneling amplitude following from the WKB approximation is given by,
\begin{eqnarray}
\Gamma\sim e^{-2{\textrm{Im}}~{\cal{S}}}=e^{\Delta S_{{\textrm{bh}}}}
\label{tunneling}
\end{eqnarray}
where the result is expressed more naturally in terms of the black hole entropy change \cite{Wilczek}. To understand the last identification ($\Gamma=e^{\Delta S_{{\textrm{bh}}}}$)
 consider a process where a black hole emits a shell of energy. We denote the initial state and final state by the levels $i$ and $f$. In thermal equilibrium,
\begin{eqnarray}
\frac{dP_i}{dt}=P_i P_{i\rightarrow f}-P_f P_{f\rightarrow i}=0
\label{detail}
\end{eqnarray}
where $P_a$ denotes the probability of getting the system in the macrostate $a (a=i,f)$ and $P_{a\rightarrow b}$ denotes the transition probability from the state $a$ to $b$ ($a,b=i,f$). According to statistical mechanics, the entropy of a given state (specified by its macrostates) is a logarithmic function of the total number of microstates ($S_{\textrm{bh}}={\textrm {log}}\Omega$). So the number of microstates $\Omega$ for a given black hole is $e^{S_{{\textrm{bh}}}}$. Since the probability of getting a system in a particular macrostate is proportional to the number of microstates available for that configuration, we get from (\ref{detail}), 
\begin{eqnarray}
e^{S_i} P_{{\textrm {emission}}}=e^{S_f}P_{{\textrm {absorption}}}
\label{detail1}
\end{eqnarray}  
where $P_{{\textrm {emission}}}$ is the emission probability $P_{i\rightarrow f}$ and $P_{{\textrm {absorption}}}$ is the absorption probability $P_{f\rightarrow i}$. So the tunneling amplitude is given by,
\begin{eqnarray}
\Gamma = \frac{P_{{\textrm {emission}}}}{P_{{\textrm {absorption}}}} = e^{S_f-S_i}=e^{\Delta S_{{\textrm{bh}}}}
\label{detail2}
\end{eqnarray}
thereby leading to the correspondance,
\begin{eqnarray}
\Delta S_{{\textrm{bh}}} = -2{\textrm{Im}}~ {\cal{S}} 
\label{1.10}
\end{eqnarray}
that follows from (\ref{tunneling}). We mention that the above relation (\ref{1.10}) has also been shown using semiclassical arguments based on the second law of thermodynamics\cite{sarkar} or on the assumption of entropy being proportional to area\cite{keski,pilling}. But such arguments are not used in our derivation. Rather our analysis has some points of similarity with the physical picture suggested in \cite{Wilczek} leading to a general validity of (\ref{detail2}).  
This implies that when quantum effects are taken into consideration, both sides of (\ref{1.10}) are modified keeping the functional relationship identical. In our analysis we will show that self consistency is preserved by (\ref{1.10}).

In order to write the black hole entropy in terms of its mass alone we have to substitute the value of $\omega$ in terms of $M$ for which the black hole is stable i. e.
\begin{eqnarray}
\frac{d(\Delta S_{{\textrm{bh}}})}{d\om}=0.
\end{eqnarray}
Using (\ref{1.5}) and (\ref{1.10}) in the above equation we get,
\begin{eqnarray}
\frac{1}{{\cal{K}}(M-\omega)}=0.
\end{eqnarray}
The roots of this equation are written in the form
\begin{eqnarray}
\om=\ph(M)
\label{17}
\end{eqnarray}
which means
\begin{eqnarray}
\frac{1}{{\cal{K}}(M-\ph(M))}=0.
\label{18}
\end{eqnarray}
This value of $\omega$ from eq. (\ref{17}) is substituted back in the expression of $\Delta S_{{\textrm{bh}}}$ to yield,
\begin{eqnarray}
\Delta S_{{\textrm{bh}}}=-2\pi\int_{0}^{\phi(M)}\frac{d\omega'}{{\cal{K}}(M-\omega')}.
\end{eqnarray}
Having obtained the form of entropy change, we are now able to give an expression of entropy for a particular state. We recall the simple definition of entropy change
\begin{eqnarray}
\Delta S_{{\textrm{bh}}}=S_{{\textrm{final}}}-S_{{\textrm{initial}}}.
\end{eqnarray}
Now setting the black hole entropy at the final state to be zero we get the expression of entropy as
\begin{eqnarray}
S_{{\textrm{bh}}}=S_{{\textrm{initial}}}=-\Delta S_{{\textrm{bh}}}=2\pi\int_{0}^{\ph(M)} \frac{d\omega'}{{\cal{K}}(M-\omega')}. 
\end{eqnarray}
From the second law of thermodynamics,
we write the inverse black hole temperature as,
\begin{eqnarray}
\frac{1}{T_h}&=&\frac{dS_{{\textrm{bh}}}}{dM}
\label{secondlaw}
\\&=&2\pi\frac{d}{dM}\int_0^{\ph(M)}\frac{d\omega'}{{\cal{K}}(M-\omega')}.\no
\end{eqnarray}
Using the identity,
\begin{eqnarray}
\frac{dF(x)}{dx}=f(x,b(x))b'(x)-f(x,a(x))a'(x)+\int_{a(x)}^{b(x)}\frac{\p}{\p x}f(x,t)dt
\end{eqnarray}
for,
\begin{eqnarray}
F(x)=\int_{a(x)}^{b(x)}f(x,t)dt
\end{eqnarray}
we find,
\begin{eqnarray}
\frac{1}{T_h}=2\pi\big[\frac{1}{{\cal{K}}(M-\ph(M))}\ph'(M)-\int_{0}^{\ph(M)}\frac{1}{[{\cal K}(M-\om')]^2}\frac{\p {\cal K}(M-\om')}{\p (M-\om')}d\om'
\big].
\label{21}
\end{eqnarray}
Making the change of variable $x=M-\om'$ in the second integral we obtain,
\begin{eqnarray}
\frac{1}{T_h}=2\pi\big[\frac{\ph'(M)-1}{{\cal{K}}(M-\ph(M))}+\frac{1}{{\cal K}(M)}\big].
\label{21xy}
\end{eqnarray}
Finally, making use of (\ref{18}), the cherished expression for the Hawking temperature follows,
\begin{eqnarray}
T_h=\frac{{\cal K}}{2\pi}=\frac{1}{4\pi}\sqrt{f'(r_h)g'(r_h)}.
\label{31}
\end{eqnarray}

For a consistency check, consider the second law of thermodynamics which is now written as,
\begin{eqnarray}
dM=d\om'=T_hdS_{{\textrm{bh}}}=\frac{{\cal K}(M)}{2\pi}dS_{{\textrm{bh}}}.
\end{eqnarray}
Inserting in (\ref{1.5}), yields,
\begin{eqnarray}
{\textrm {Im}}~{\cal S}=\frac{1}{2}\int_{S_{{\textrm{bh}}}(M)}^{S_{{\textrm{bh}}}(M-\om)}dS_{{\textrm{bh}}}=-\frac{1}{2}\Delta S_{{\textrm {bh}}}
\end{eqnarray}
thereby reproducing (\ref{1.10}). This shows the internal consistency of the tunneling approach.
\section{Schwarzschild black hole in noncommutative space}
We shall use the general formulation developed in the previous section for a noncommutative Schwarzschild black hole. The appropriate metric will be constructed from which the horizon of the black hole will be defined. The thermodynamic quantities of this black hole will be calculated in the next section.

In a commutative space, the mass density of a point particle is expressed as a product of its mass with the Dirac delta function. But in a noncommutative space, such a description of point mass is not possible due to the fuzziness of space which arises as a consequence of position-position uncertainty relation. We denote the noncommutative parameter by $\theta$ which is considered to be a small ($\sim{\textrm{Planck length}}^2$) positive number. Now to introduce the noncommutative correction in the expression of mass density we replace the Dirac delta function by a Gaussian distribution of minimal width $\sqrt{\theta}$ i. e.\cite{smail,park}
 \begin{eqnarray}
\rho_\theta=\frac{M}{(4\pi\theta)^{\frac{3}{2}}}e^{-\frac{r^2}{4\theta}}.
\label{2.2}
\end{eqnarray}
So the mass is no longer located at a point, instead it is smeared around a region $\sqrt{\theta}$. Therefore the mass of the black hole can be determined by integrating (\ref{2.2}) over a volume of radius $r$. This is found to be,
\begin{eqnarray}
m_\theta(r)=\int_0^r 4\pi r'^2 \rho_{\theta}(r')dr' = \frac{2M}{\sqrt{\pi}}\gamma(\frac{3}{2},\frac{r^2}{4\theta}).
\label{2.4}
\end{eqnarray}
where $\gamma(\frac{3}{2},\frac{r^2}{4\theta})$ is the lower incomplete gamma function which is discussed in the Appendix. So in the $\theta\rightarrow 0$ limit, the incomplete $\gamma$ function becomes the usual gamma function ($\Gamma_{\textrm{total}}$) and $m_\theta(r)\rightarrow M$ which is the commutative limit of the noncommutative mass $m_\theta(r)$. Substituting this in the mass term of the Schwarzschild space time
\begin{eqnarray}
ds^2 = -(1-\frac{2M}{r})dt^2 + (1-\frac{2M}{r})^{-1}dr^2 + r^2 d\Omega^2
\label{2.1}
\end{eqnarray}
we get the noncommutative Schwarzschild metric,
\begin{eqnarray}
ds^2 = -\Big(1-\frac{4M}{r\sqrt{\pi}}\gamma(\frac{3}{2},\frac{r^2}{4\theta})\Big)dt^2 + \Big(1-\frac{4M}{r\sqrt{\pi}}\gamma(\frac{3}{2},\frac{r^2}{4\theta})\Big)^{-1}dr^2 + r^2 d\Omega^2.
\label{2.5}
\end{eqnarray}
The same line element is also obtained by solving Einstein's equation with (\ref{2.2}) as the matter source\cite{smail}. The event horizon can be found where $g^{rr}(r_h)=0$, that is
\begin{eqnarray}
r_h = \frac{4M}{\sqrt{\pi}}\gamma\Big(\frac{3}{2},\frac{r_h^2}{4\theta}\Big).
\label{2.6}
\end{eqnarray}
This equation cannot be solved for $r_h$ in a closed form. In the large radius regime ($\frac{r_h^2}{4\theta}>>1$) we use the expanded form of the incomplete $\gamma$ function given in the Appendix (eq. (\ref{app4})) to solve eq. (\ref{2.6}) by iteration. Keeping upto the order $\frac{1}{\sqrt{\theta}}e^{-\frac{M^2}{\theta}}$, we find
\begin{eqnarray}
r_h\simeq2M\Big(1-\frac{2M}{\sqrt{\pi\theta}}e^{-\frac{M^2}{\theta}}\Big).
\label{2.7}
\end{eqnarray}
It might be mentioned that there are other approaches \cite{Lopez,chai,mukherjee,kobak} of introducing noncommutativity in curved space time metric. Contrary to the present approach, however, there the metric is not spherically symmetric. The relevance of this criterion in the present analysis stems from the fact that thermodynamic properties of the noncommutative black holes, analysed in the next section, use the results of section 2 which are based on a static spherically symmetric metric.

\section{Noncommutative Hawking temperature, tunneling rate and entropy in the presence of back reaction}
We take the units $G=c=k_{\textrm B}=1$, in which{\footnote {Planck  length  $l_P =(\hbar G/c^3)^{1/2}$, Planck  mass  $M_P =(\hbar c/G)^{1/2}$}} Planck length $l_{\textrm p}$=Planck mass $M_{\textrm p}$$=\sqrt{\hbar}$. Since a loop expansion is equivalent to an expansion in powers of the Planck constant, the one loop back reaction effect in the surface gravity is written as,
\begin{eqnarray}
{\cal{K}} = {\cal{K}}_0(r_h)+\xi{\cal{K}}_0(r_h)
\label{xi}
\end{eqnarray}
where ${\cal{K}}_0$ is the noncommutative classical surface gravity at the horizon of the black hole and $\xi$ is a dimensionless constant having magnitude of the order ${\hbar}$. From dimensional arguments, therefore, it has the structure, 
\begin{eqnarray}
\xi=\beta\frac{M_{\textrm p}^2}{m^2_\theta}
\end{eqnarray}
where $\beta$ is a pure numerical factor. In the commutative picture $\beta$ is known to be related to the trace anomaly coefficient\cite{Lousto,Fursaev}. Putting this form of $\xi$ in (\ref{xi}) we get,
\begin{eqnarray}
{\cal{K}} = {\cal{K}}_0(r_h)\Big(1+\beta\frac{M_{\textrm p}^2}{m^2_\theta}\Big).
\label{4.1z}
\end{eqnarray}
A similar expression was obtained earlier in \cite{York,Lousto} for the commutative case. Eq. (\ref{4.1z}) is recast as,
\begin{eqnarray}
{\cal{K}} = {\cal{K}}_0(r_h)\Big(1+\frac{\alpha}{m^2_\theta(r_h)}\Big)
\label{4.1}
\end{eqnarray}
where $\alpha=\beta M_{\textrm p}^2$. Since as mentioned already, the noncommutative parameter $\theta$ is of the order of $l_{\textrm p}^2$, $\alpha$ and $\theta$ are of the same order. This fact will be used later when doing the graphical analysis.

In order to calculate the right hand side of (\ref{4.1}), we need to obtain an expression for noncommutative classical surface gravity at the horizon of the black hole $({\cal{K}}_0(r_h))$. This is done by using (\ref{1.06}). For the classical noncommutative Schwarzschild spacetime 
\begin{eqnarray}
f(r)=g(r)=  1-\frac{4M}{r\sqrt{\pi}}\gamma(\frac{3}{2},\frac{r^2}{4\theta}).
\label{4.2}
\end{eqnarray}
The value of ${\cal{K}}_0(r_h)$ is thus found to be,
\begin{eqnarray}
{\cal{K}}_0(r_h) = \frac{f'(r_h)}{2}= \frac{1}{2}\Big[\frac{1}{r_h}-\frac{r^2_h}{4\theta^{\frac{3}{2}}} \frac{e^{-\frac{r^2_h}{4\theta}}}{\gamma\Big(\frac{3}{2},\frac{r^2_h}{4\theta}\Big)}\Big].
\label{4.3}
\end{eqnarray}
Inserting (\ref{4.3}) in (\ref{4.1}) we get,
\begin{eqnarray}
{\cal{K}} = \frac{1}{2}\Big[\frac{1}{r_h}-\frac{r^2_h}{4\theta^{\frac{3}{2}}} \frac{e^{-\frac{r^2_h}{4\theta}}}{\gamma\Big(\frac{3}{2},\frac{r^2_h}{4\theta}\Big)}\Big]\Big(1+\frac{\alpha}{m^2_\theta(r_h)}\Big).
\label{4.1mn}
\end{eqnarray}
In order to write the above equation completely in terms of $r_h$ we have to express the mass $m_\theta$ in terms of $r_h$. For that we compare eqs. (\ref{2.4}) and (\ref{2.6}) to get,
\begin{eqnarray}
m_{\theta}(r_h)=\frac{r_h}{2}.
\label{4.4}
\end{eqnarray}
This relation is the noncommutative deformation of the standard radius-mass relation for the usual (commutative space) Schwarzschild black hole. Expectedly in the limit $\theta\rightarrow 0$ eq. (\ref{4.4}) reduces to its commutative version $r_h=2M.$


Substituting (\ref{4.4}) in (\ref{4.1mn}) we get the value of modified noncommutative surface gravity
\begin{eqnarray}
{\cal{K}} = \frac{1}{2}\Big[\frac{1}{r_h}-\frac{r^2_h}{4\theta^{\frac{3}{2}}} \frac{e^{-\frac{r^2_h}{4\theta}}}{\gamma\Big(\frac{3}{2},\frac{r^2_h}{4\theta}\Big)}\Big]\Big(1+\frac{4\alpha}{r^2_h}\Big).
\label{4.5}
\end{eqnarray} 
So from (\ref{31}), the modified noncommutative Hawking temperature including the effect of back reaction is given by,
\begin{eqnarray}
T_h=\frac{{\cal{K}}}{2\pi}=\frac{1}{4\pi}\Big[\frac{1}{r_h}-\frac{r^2_h}{4\theta^{\frac{3}{2}}} \frac{e^{-\frac{r^2_h}{4\theta}}}{\gamma\Big(\frac{3}{2},\frac{r^2_h}{4\theta}\Big)}\Big]\Big(1+\frac{4\alpha}{r^2_h}\Big).
\label{4.6}
\end{eqnarray}
If the back reaction is ignored (i. e. $\alpha =0$), the expression for the Hawking temperature agrees with that given in \cite{smail}. Also for the $\theta\rightarrow 0$ limit, one can recover the standard result \cite{Fursaev,majhi} for Hawking temperature (with back reaction)
\begin{eqnarray}
T_h=T_H(1+\frac{\alpha}{M^2})
\label{4.61}
\end{eqnarray}
where $T_H=\frac{1}{8\pi M}$ is the semiclassical Hawking temperature for the Schwarzschild black hole.

\begin{figure}[t] 
\centering
\includegraphics[angle=0,width=15cm,height=15cm]{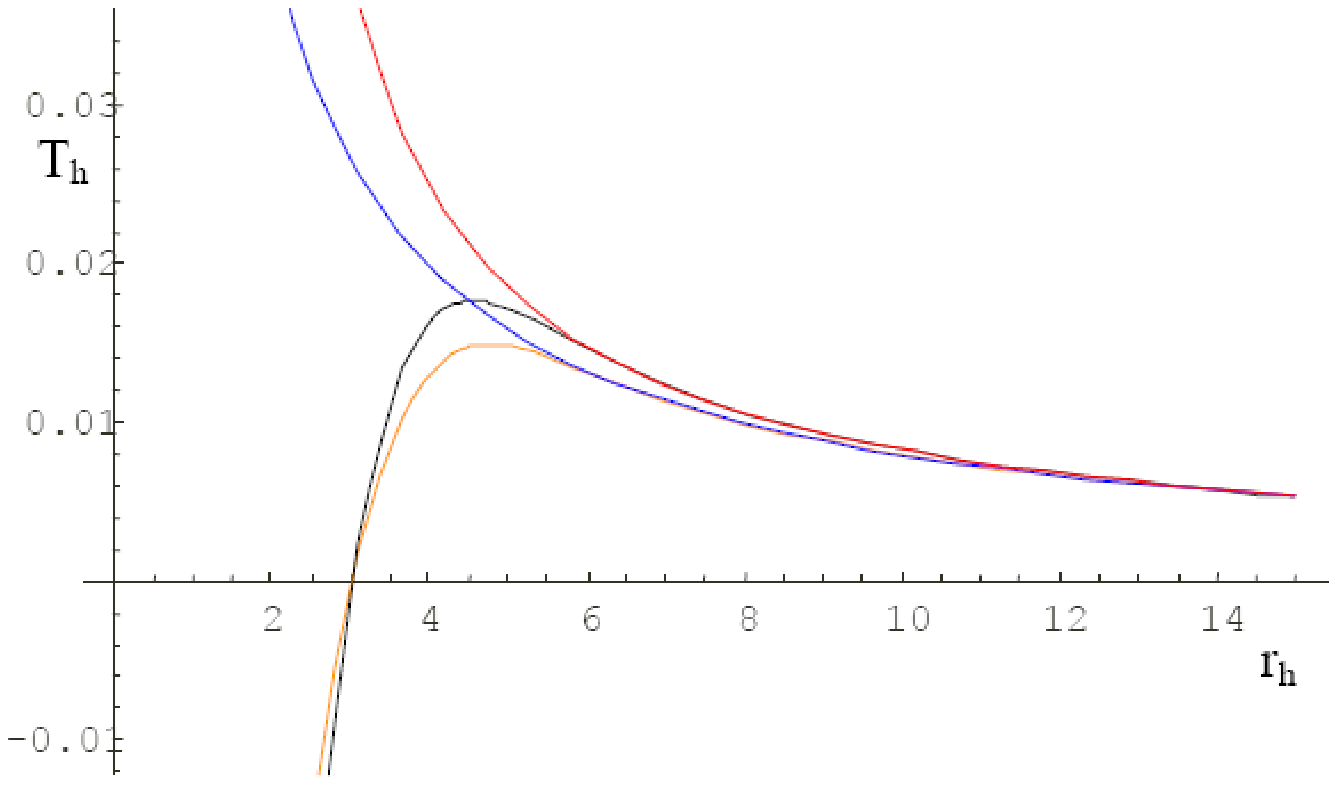}
\caption[]{\it{{$T_h$ Vs. $r_h$ plot (Here $\alpha=\theta$, $\alpha$ and $\theta$ are positive).\\ 
$r_h$ is plotted in units of $\sqrt\theta$ and $T_h$ is plotted in units of $\frac{1}{\sqrt\theta}$.\\
Red curve: $\alpha\neq 0, \theta=0$.\\
Blue curve: $\alpha= 0, \theta= 0$.\\
Black curve: $\alpha\neq 0, \theta\neq 0$.\\
Yellow curve: $\alpha= 0, \theta\neq 0$.}}}
\label{fig1}
\end{figure}

\begin{figure}[t] 
\centering
\includegraphics[angle=0,width=15cm,height=15cm]{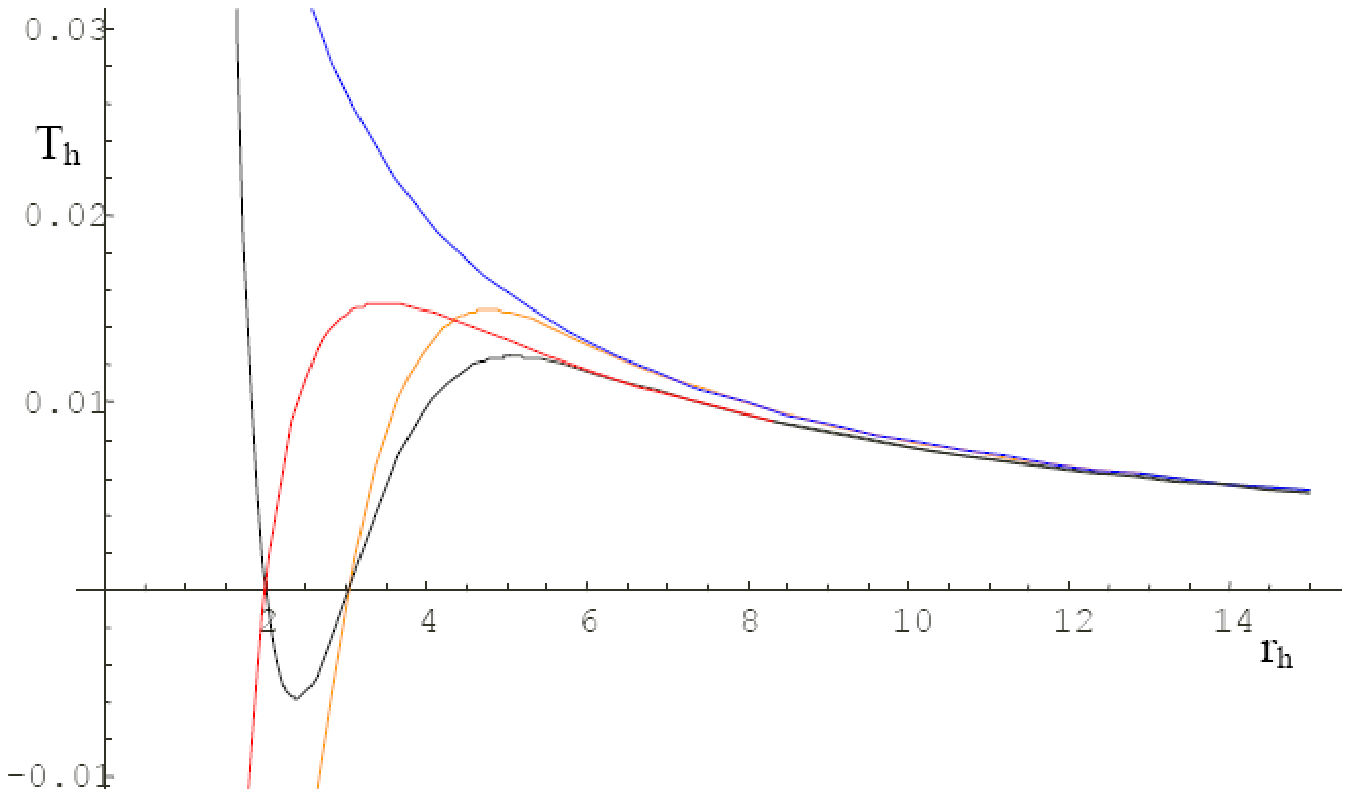}
\caption[]{{\it{$T_h$ Vs. $r_h$ plot (Here $|\alpha|=\theta$, $\alpha$ is negative but $\theta$ is positive).\\
$r_h$ is plotted in units of $\sqrt\theta$ and $T_h$ is plotted in units of $\frac{1}{\sqrt\theta}$.\\
Red curve: $\alpha\neq 0, \theta=0$.\\
Blue curve: $\alpha= 0, \theta= 0$.\\
Black curve: $\alpha\neq 0, \theta\neq 0$.\\
Yellow curve: $\alpha= 0, \theta\neq 0$.}}}
\label{fig2}
\end{figure}

   In the standard (commutative) case $T_h$ diverges as $M\rightarrow 0$ and this puts a limit on the validity of the conventional description of Hawking radiation. Against this scenario, temperature (\ref{4.6}) includes noncommutative and back reaction effects which are relevant at distances comparable to $\sqrt\theta$. The behaviour of the temperature $T_h$ as a function of horizon radius $r_h$ is plotted in fig.(\ref{fig1}) (with positive $\alpha$) and in fig.(\ref{fig2}) (with negative $\alpha$).

Fig.(1) shows that in the region $r_h\simeq\sqrt\theta$, the effect of noncommutativity significantly changes the nature of commutative space curves. Interestingly two noncommutative curves, whether including back reaction or not are qualitatively same. Both of them attain a maximum value at $r_h=r_0\simeq 4.7\sqrt\theta$ and then sharply drop to zero forming an extremal black hole. In the region $r_h<r_0$ there is no black hole, because physically $T_h$ cannot be negative. The only difference between them is that the back reaction effect increases the maximum temperature by $20\%$. Infact, in the commutative space also, back reaction effect increases the value of Hawking temperature. But quite contrary to the noncommutative curves, both of them diverge as $r_h\rightarrow 0$. As easily observed, the Hawking paradox is circumvented by noncommutativity, with or without back reaction. This was also noted in \cite{smail} where, however, the quantitative effects of back reaction were not considered.  

On the other hand fig.(2) shows that if any of the two effects (i.e. either noncommutativity or back reaction) is present $T_h$ drops to zero. For $\alpha=0, \theta\neq 0$ (yellow curve) $T_h$ becomes zero at $r_h=r_0\simeq 3.0\sqrt\theta$ and for $\alpha\neq 0, \theta=0$ (red curve) it becomes zero at $r_h=r_0\simeq 2.0\sqrt\theta$ . These cases therefore bypass the Hawking paradox. But for noncommutative black hole with back reaction ($\alpha\neq0,\theta\neq0$), $T_h$ is zero for two values of $r_h$: $r_h\simeq 3.0\sqrt\theta$ and $r_h=2.0\sqrt\theta$ and then it diverges towards positive infinity. This is not physically possible since after entering the forbidden zone it resurfaces on the allowed sector. So for both noncommutativity and back reaction effect, $\alpha$ can never be negative.

Having obtained the Hawking temperature of the black hole we calculate the Bekenstein-Hawking entropy. The expression of entropy can be obtained from the second law of thermodynamics. But instead of using it we employ the formula (\ref{1.10}) to calculate the entropy. Using (\ref{2.7}) the modified surface gravity (\ref{4.5}) can be approximately expressed in terms of $M$. To the leading order, we obtain, 
\begin{eqnarray}
{\cal{K}}(M)&=&\frac{M^2+\alpha}{4M^3}\Big[1-\frac{4M^5}{(M^2+\alpha)\theta\sqrt{\pi\theta}}e^{-\frac{M^2}{\theta}}\Big]+{\cal{O}}(\frac{1}{\sqrt\theta}e^{-\frac{M^2}{\theta}}).
\label{4.7}
\end{eqnarray}
Substituting this in (\ref{1.5}) and then integrating over $\omega'$ we have, 
\begin{eqnarray}
{\textrm{Im}}~{\cal{S}} &=& 4\pi\omega(M-\frac{\omega}{2})+2\pi\alpha \ln{\Big[\frac{(M-\omega)^2+\alpha}{M^2+\alpha}\Big]}- 8\sqrt{\frac{\pi}{\theta}}M^3 e^{-\frac{M^2}{\theta}}
\nonumber
\\
&+& 8\sqrt{\frac{\pi}{\theta}}(M-\omega)^3 e^{-\frac{(M-\omega)^2}{\theta}}+\textrm{const.(independent of $M$)}+{\cal O}(\sqrt\theta e^{-\frac{M^2}{\theta}}).
\label{4.8}
\end{eqnarray}
So by the relation (\ref{tunneling}) the modified tunneling probability due to noncommutativity and back reaction effects is, 
\begin{eqnarray}
\Gamma&\sim& \Big[1-\frac{2\omega(M-\frac{\omega}{2})}{M^2+\alpha}\Big]^{-4\pi\alpha} \textrm{exp}\Big[16\sqrt{\frac{\pi}{\theta}}M^3e^{-\frac{M^2}{\theta}}-16\sqrt{\frac{\pi}{\theta}}(M-\omega)^3e^{-\frac{(M-\omega)^2}{\theta}}\nonumber
\\
&+&\textrm{const.(independent of $M$)}\Big]
\textrm{exp}\Big[-8\pi\omega(M-\frac{\omega}{2})\Big]. 
\label{4.9}
\end{eqnarray}
The last exponential factor of the tunneling probability was previously obtained by Parikh and Wilczek \cite{Wilczek} where neither noncommutativity nor back reaction effects were considered. The factors before this exponential are actually due the effect of back reaction and noncommutativity. It will eventually give the correction to the Bekenstein-Hawking entropy and the Hawking temperature as will be shown below. Taking $\theta\rightarrow 0$ limit we can immediately reproduce the commutative tunneling rate for Schwarzschild black hole with back reaction effect\cite{majhi}.

   We are now in a position to obtain the noncommutative deformation of the Bekenstein-Hawking area law. The first step is to compute the entropy change $\Delta S_{{\textrm{bh}}}$. Using (\ref{tunneling}) and (\ref{4.9}) we obtain, to the leading order,
\begin{eqnarray}
\Delta S_{{\textrm{bh}}}&\simeq& -8\pi\omega(M-\frac{\omega}{2})-4\pi\alpha \ln{\Big[\frac{(M-\omega)^2+\alpha}{M^2+\alpha}\Big]}+16 \sqrt{\frac{\pi}{\theta}}M^3 e^{-\frac{M^2}{\theta}}
\nonumber
\\
&-& 16\sqrt{\frac{\pi}{\theta}}(M-\omega)^3 e^{-\frac{(M-\omega)^2}{\theta}}+\textrm{const.(independent of $M$)}.
\label{4.10}
\end{eqnarray}
Next using the stability criterion $\frac{d(\Delta S_{{\textrm{bh}}})}{d\omega}=0 $ for the black hole, one obtains the only physically possible solution for $\omega$ as
 $\omega=M$. Substituting this value of $\omega$ in (\ref{4.10}) and setting $S_{{\textrm{final}}}=0$ we have the Bekenstein-Hawking entropy
\begin{eqnarray}
S_{{\textrm{bh}}}=S_{{\textrm{initial}}}&\simeq& 4\pi M^2-4\pi\alpha\ln{(\frac{M^2}{\alpha}+1)}
\nonumber
\\
&-& 16\sqrt{\frac{\pi}{\theta}}M^3 e^{-\frac{M^2}{\theta}}+\textrm{const.(independent of $M$)}.
\label{4.13}
\end{eqnarray}
Neglecting the back reaction effect ($\alpha=0$) the above expression of black hole entropy is written as 
\begin{eqnarray}
S_{{\textrm{bh}}}\simeq 4\pi M^2- 16\sqrt{\frac{\pi}{\theta}}M^3 e^{-\frac{M^2}{\theta}}.
\label{en1}
\end{eqnarray}
Now in order to write the above equation in terms of the noncommutative horizon area ($A$), we use (\ref{2.7}) to obtain,
\begin{eqnarray}
A=4\pi r_h^2=16\pi M^2 - 64 \sqrt{\frac{\pi}{\theta}} M^3 e^{-\frac{M^2}{\theta}}+{\cal{O}}(\sqrt\theta e^{-\frac{M^2}{\theta}}).
\label{en2}
\end{eqnarray}
Comparing equations (\ref{en1}) and (\ref{en2}) we see that at the leading order the noncommutative black hole entropy satisfies the area law
\begin{eqnarray}
S_{{\textrm{bh}}}=\frac{A}{4}.
\label{en3}
\end{eqnarray}
This is functionally identical to the Bekenstein-Hawking area law in the commutative space. 

Considering $\theta\rightarrow 0$ limit in (\ref{4.13}) we have the corrected form of Bekenstein-Hawking entropy for commutative Schwarzschild black hole with back reaction effect \cite{Fursaev,majhi}. The well known logarithmic correction \cite{Page} is reproduced.

     Now using the second law of thermodynamics (\ref{secondlaw}) we can find the corrected form of the Hawking temperature $T_h$ due to back reaction. This is obtained from (\ref{4.13}) as,
\begin{eqnarray}
\frac{1}{T_h}=\frac{dS_{{\textrm{bh}}}}{dM}=\frac{8\pi M^3}{M^2+\alpha}+32\frac{\sqrt{\pi}}{\theta^{\frac{3}{2}}}M^4 e^{-\frac{M^2}{\theta}}+{\cal{O}}(\frac{1}{\sqrt\theta}e^{-\frac{M^2}{\theta}}).
\label{4.14}
\end{eqnarray}
Therefore the corrected noncommutative Hawking temperature is given by
\begin{eqnarray}
T_h &=& \frac{M^2+\alpha}{8\pi M^3} -\frac{M^2}{2(\pi\theta)^{\frac{3}{2}}}e^{-\frac{M^2}{\theta}}+ {\cal{O}}(\frac{1}{\sqrt\theta}e^{-\frac{M^2}{\theta}}).
\label{1.17}
\end{eqnarray}

We now provide a simple consistency check on the relation (\ref{4.6}). The Hawking temperature is recalculated using this relation and showing that it reproduces (\ref{1.17}). For the large radius limit, (\ref{4.6}) is approximately written as,
\begin{eqnarray}
T_h\simeq\frac{1}{4\pi}\big[\frac{1}{r_h}-\frac{r_h^2}{2\sqrt{\pi}\theta^{3/2}}e^{-\frac{r_h^2}{4\theta}}\big]\big(1+\frac{4\alpha}{r_h^2}\big).
\label{neweqn}
\end{eqnarray}
Now the approximated form of $r_h$ in terms of $M$ (\ref{2.7}) is substituted in (\ref{neweqn}) to get the relation (\ref{1.17}) upto the leading order in the noncommutative parameter. This shows the self consistency of our calculation.

For $\alpha=\theta=0$, the expression (\ref{1.17}) reduces to the usual Hawking temperature $T_H=\frac{1}{8\pi M}$ for a Schwarzschild black hole. Also, keeping the back reaction ($\alpha$) but taking $\theta\rightarrow 0$ limit, we reproduce the commutative Hawking temperature\cite{Fursaev,majhi}.

\section{Conclusions}
We have given a completely general derivation of Hawking temperature in terms of the surface gravity by considering the action of an outgoing particle crossing the black hole horizon due to quantum mechanical tunneling. The expression of temperature was known long before\cite{Bardeen,1,3,2} from a comparison between two classical laws. One is the law of black hole thermodynamics which states that the mass change is proportional to the change of horizon area multiplied by surface gravity at the horizon. The other is the area law according to which the black hole entropy is proportional to the surface area of the horizon. The important point of our derivation is that it is not based on either of these two classical laws.

The other significant point of this paper is the application of our formulation to a noncommutative Schwarzschild metric, keeping in mind the consequence of back reaction. Several thermodynamic entities like the temperature and entropy are computed. The tunneling rate is also derived. The temperature, in particular, is obtained in a closed form. This result is analyzed in detail using two graphical representations. We give particular attention to the small scale behaviour of black hole temperature where the effects of both noncommutativity and back reaction are highly nontrivial. The graphs presented here are naturally more general than \cite{smail} and \cite{Lousto}, because in \cite{smail} the effect of back reaction was not included and in \cite{Lousto} space time was taken to be commutative in nature. Expectedly in suitable limits, the results of our paper reduce to that of \cite{smail} and \cite{Lousto} but the combination of noncommutativity and back reaction, as shown here, gives new results at small scale. In particular, it is shown that in the presence of both noncommutativity and back reaction, the back reaction parameter $\alpha$ cannot be negative. Interestingly, even for the commutative case, arguments based on quantum geometry \cite{kaul,Page,majhi} fix a positive value for $\alpha$.

In the noncommutative analysis, with positive $\alpha$, (Fig 1), the maximum Hawking temperature gets enhanced in the presence of back reaction. However, the Hawking paradox is avoided whether or not the back reaction is included.

   Apart from the temperature, other variables like the tunneling rate and entropy are given upto the leading order in the noncommutative parameter. The entropy is expressed in terms of the area. The result is a noncommutative deformation of the Bekenstein-Hawking area law, preserving the usual functional form. Since both $T_h=\frac{{\cal{K}}}{2\pi}$ and the area law retain their standard forms it suggests that the laws of noncommutative black hole thermodynamics are a noncommutative deformation of the usual laws.

   As a final remark we mention that although our results are presented for the noncommutative Schwarzschild metric, the formulation is resilient enough to discuss other types of noncommutative black holes.

\section*{Appendix}

{\bf{Incomplete gamma function:}}
The lower incomplete gamma function is given by
\begin{eqnarray}
\gamma(a,x)=\int_0^x t^{a-1}e^{-t}dt
\label{app1}
\end{eqnarray}
whereas the upper incomplete gamma function is
\begin{eqnarray}
\Gamma(a,x)=\int_x^\infty t^{a-1}e^{-t}dt
\label{app2}
\end{eqnarray}
and they are related to the total gamma function through the following relation
\begin{eqnarray}
\Gamma_{{\textrm{total}}}(a)=\gamma(a,x)+\Gamma(a,x)=\int_0^\infty t^{a-1}e^{-t}dt.
\label{app3}
\end{eqnarray}
Furthermore, for large $x$, i.e. $x>>1$, the asymptotic expansion of the lower incomplete gamma function is given by
\begin{eqnarray}
\gamma(\frac{3}{2},x)&=&\Gamma_{{\textrm{total}}}(\frac{3}{2})-\Gamma(\frac{3}{2},x)
\nonumber
\\
&\simeq&\frac{\sqrt{\pi}}{2}\Big[1-e^{-x}\sum_{p=0}^{\infty}\frac{x^{\frac{1-2p}{2}}}{\Gamma_{{\textrm{total}}}(\frac{3}{2}-p)}\Big].
\label{app4}
\end{eqnarray}
Using the definition (\ref{app1}) and then integrating by parts we have
\begin{eqnarray}
\gamma(a+1,x)=\int_0^x t^ae^{-t}dt&=&-t^ae^{-t}|_0^x+a\int_o^x t^{a-1}e^{-t}dt
\nonumber
\\
&=&-x^ae^{-x}+a\gamma(a,x).
\label{app7}
\end{eqnarray}
Similarly by the definition (\ref{app2}) one can show
\begin{eqnarray}
\Gamma(a+1,x)=x^ae^{-x}+a\Gamma(a,x).
\label{app8}
\end{eqnarray}

{\bf{Some useful Formulas:}}

\begin{eqnarray}
I_1=\int_a^b e^{-\alpha x^2}dx=\frac{1}{2{\alpha}^{\frac{1}{2}}}\Big[\sqrt{\pi}-\gamma(\frac{1}{2},\alpha a^2)-\Gamma(\frac{1}{2},\alpha b^2)\Big]
\label{app51}
\end{eqnarray}
\begin{eqnarray}
I_2=\int_a^b x^2e^{-\alpha x^2}dx=\frac{1}{2{\alpha}^{\frac{3}{2}}}\Big[\frac{\sqrt{\pi}}{2}-\gamma(\frac{3}{2},\alpha a^2)-\Gamma(\frac{3}{2},\alpha b^2)\Big]
\label{app5}
\end{eqnarray}
\begin{eqnarray}
I_3=\int_a^b x^4e^{-\alpha x^2}dx=\frac{1}{2{\alpha}^{\frac{5}{2}}}\Big[\frac{3\sqrt{\pi}}{4}-\gamma(\frac{5}{2},\alpha a^2)-\Gamma(\frac{5}{2},\alpha b^2)\Big]
\label{app6}
\end{eqnarray}

\end{document}